\documentclass[reprint,prd,onecolumn,notitlepage,nofootinbib,11pt]{revtex4-1}
\usepackage{amsmath,amssymb,graphicx,booktabs,float}
\usepackage[colorlinks=true,citecolor=blue,linkcolor=blue,urlcolor=blue]{hyperref} 
\usepackage[font={small},flushleft,indent]{caption}
\usepackage{subcaption}

\begin{document}

\title{Correspondence between multileaf topology of closed geodesics and spatiotemporal autocorrelations of hotspot images in Schwarzschild spacetime}
\author{Fengting Xie}
\email{xiefengting@stu.cqu.edu.cn}
\author{Qing-Hua Zhu}
\email[Corresponding author: ]{zhuqh@cqu.edu.cn}
\author{Xin Li}
\email[Corresponding author: ]{lixin1981@cqu.edu.cn}
\affiliation{Department of Physics and Chongqing Key Laboratory for Strongly Coupled Physics, Chongqing University, Chongqing 401331, China}

\begin{abstract}
The classification of relativistic closed orbits by their multileaf structures provides a framework for studying strong-field dynamics. 
Identifying these structures in astronomical images remains challenging. 
Using ray tracing, we construct the spatiotemporal autocorrelations of the primary image of a pointlike hotspot moving along bound closed geodesics around a Schwarzschild black hole. 
These orbits are classified by three integers $(z,w,v)$, where $z$ counts the leaves, $w$ counts the additional whirls during each radial period, and $v$ specifies the order in which the orbital leaves are traced.
For the orbit families examined, our numerical results establish a correspondence between the topological integers $(z,w,v)$ and the numbers of correlation bands $N_{\rm band}$ and recurrence points $N_{\rm rec}$.
The integers are recovered as $z=N_{\rm rec}-1$, $w=\left\lfloor (N_{\rm band}+1)/(N_{\rm rec}-1)\right\rfloor-1$, and $v=(N_{\rm band}+1)\bmod(N_{\rm rec}-1)$.
Here, $\lfloor x\rfloor$ denotes the greatest integer not exceeding $x$, and $\bmod$ denotes the remainder operation.
These relations provide a quantitative method for recovering closed-orbit topology from hotspot image autocorrelations.
\end{abstract}
\maketitle

\section{Introduction}

Orbital motion near black holes and its associated radiation provide important avenues for probing strong-field dynamics and spacetime geometry~\cite{Ryan:1995wh,Glampedakis:2002ya,Bambi:2015kza,Cardenas-Avendano:2024mqp}.
The classification of periodic orbits provides a systematic framework for organizing bound orbital dynamics and understanding the structure of more general, nonperiodic motion~\cite{Levin:2008mq,Misra:2010pu,Grossman:2012,Lim:2024mkb}.
In the strong-field regime, bound eccentric orbits can exhibit distinctive multileaf structures and zoom–whirl behavior, revealing relativistic dynamics beyond the familiar picture of weakly precessing ellipses~\cite{Levin:2008mq}.
Closed orbits can be classified by three integers $(z,w,v)$, which specify the number of leaves, the number of additional whirls near the periastron during each radial period, and the order in which successive apastron vertices are visited, respectively~\cite{Levin:2008mq}. This classification relates orbital structures to their energies and angular momenta~\cite{Levin:2009,Grossman:2012} and has been applied to investigate orbital properties in a variety of black-hole and wormhole spacetimes~\cite{Liu:2018vea,Wei:2019zdf,Deng:2020hxw,Zhou:2020zys}.
Beyond the orbital dynamics themselves, recent studies have explored the gravitational-wave signatures associated with periodic orbits in extreme-mass-ratio systems~\cite{Tu:2023xab,Li:2024tld,Meng:2024cnq,Wang:2025hla,Chen:2025aqh,Lu:2025xlp,Gong:2025mne,Huang:2026oga}.

Observations with the Event Horizon Telescope (EHT) and GRAVITY have opened new opportunities to study the immediate environments of supermassive black holes, through horizon-scale imaging and measurements of near-infrared flare motions, respectively~\cite{EventHorizonTelescope:2019dse,EventHorizonTelescope:2019ggy,EventHorizonTelescope:2022wkp,EventHorizonTelescope:2022xqj,Abuter:2018uum,GRAVITY:2023data}. Studies of hotspots moving along timelike geodesics, including closed and quasi-periodic trajectories, have examined how their motion is reflected in images and light curves~\cite{Huang:2024wpj,Tan:2026yjg}. However, the observed images and light curves of orbiting sources depend on both their motion and relativistic effects, including gravitational lensing, Doppler shifts, and gravitational redshift~\cite{Huang:2024wpj,Zhou:2024dbc}. These effects can complicate the direct identification of multileaf orbital structures, motivating us to explore whether these orbital features can be identified indirectly.

Correlation analysis provides a useful way to extract information about source dynamics from radiation signals and has broad applications in astronomy, including studies of stellar rotation and accretion variability~\cite{Kazanas:1999hg,Berkley:2000mp,McQuillan_2014,Peterson:2004nu,Gandhi:2008qr}.
In black hole systems, correlations associated with accretion flows and gravitationally lensed images have been investigated as probes of source dynamics and the underlying spacetime geometry
\cite{Conroy:2023kec,Fukumura_2010,Chesler:2020gtw,Hadar:2020fda,Chen:2022kzv,Qian:2021aju,Emami:2023sus,Hadar:2023kau,Cardenas-Avendano:2024sgy,Zhu:2023omf,Conroy:2025nev,Harikesh:2025nmt}.
For orbiting hotspots, previous studies have investigated bandlike autocorrelation structures and correlations between different image orders~\cite{Zhu:2025jqh,Zhang:2025vyx}.
These developments suggest that correlation analysis may also help identify orbital features that are difficult to recognize directly in images.
Here, we investigate whether the spatiotemporal autocorrelation of the primary image of a hotspot moving along a closed multileaf orbit can reveal its orbital topology, characterized by $(z,w,v)$.

In this work, we consider a pointlike hotspot moving along a bound closed geodesic in the equatorial plane of a Schwarzschild black hole.
We use the ray-tracing and correlation methods developed in Refs.~\cite{Zhu:2024vxw,Zhu:2025jqh} to obtain the apparent positions and arrival times of the primary image and to construct its spatiotemporal autocorrelation.
Our numerical results show that, within one orbital period, the number of correlation bands depends on a combination of all three integers $(z,w,v)$, whereas the number of recurrence points directly determines the leaf number $z$.
These two signatures allow us to recover the three integers $(z,w,v)$ for the orbit families examined. We also investigate the effect of orbital eccentricity, which changes the correlation-band widths while preserving these counting relations in the cases considered.

The rest of the paper is organized as follows. In Sec.~\ref{sec:methods}, we introduce the closed hotspot orbits and define the spatiotemporal correlations. In Sec.~\ref{sec:topology_correlations}, we examine the correspondence between closed-orbit topology and the structure of the primary-image autocorrelation. 
In Sec.~\ref{sec:effect_ecc}, we examine the effect of orbital eccentricity on the correlation-band widths.
Finally, Sec.~\ref{sec:conclusions} summarizes the main results and discusses their implications. The appendices present correlations involving higher-order images and additional checks of the counting relations.

\section{Closed orbits and correlation pattern}\label{sec:methods}

In this section, we introduce the orbital model and correlation functions used in the subsequent parts.
We first describe bound closed geodesics in Schwarzschild spacetime and the parameters characterizing their topology and eccentricity. We then define the spatiotemporal correlations of the hotspot images, with the subsequent analysis focusing on the primary-image autocorrelation. Throughout this work, we adopt geometrized units, $G=c=1$, and measure lengths and times in units of the mass $M$ of a Schwarzschild black hole.

\subsection{Closed geodesic orbits of a hotspot}

The hotspot is modeled as a localized bright region on the accretion disk \cite{Broderick:2005}, which has been observed to undergo orbital motion around the black hole \cite{Abuter:2018uum,GRAVITY:2023data}.
In this study, we thus idealize the hotspot as a point-like source moving along a bound, noncircular timelike geodesic in the equatorial plane. The background geometry is described by the Schwarzschild line element
\begin{equation}
    \mathrm{d}s^2=-f(r)\,\mathrm{d}t^2+\frac{\mathrm{d}r^2}{f(r)}+r^2\left(\mathrm{d}\theta^2+\sin^2\theta \mathrm{d}\phi^2\right),
\end{equation}
where $f(r)=1-\frac{2M}{r}$. For equatorial motion, $\theta=\pi/2$, the geodesic equations follow from the Lagrangian
\begin{equation}
    \mathcal{L}=\frac12\left[-f(r)\left(\frac{\mathrm{d}t}{\mathrm{d}\tau}\right)^2+\frac{1}{f(r)}\left(\frac{\mathrm{d}r}{\mathrm{d}\tau}\right)^2+r^2\left(\frac{\mathrm{d}\phi}{\mathrm{d}\tau}\right)^2\right],
\end{equation}
where $\tau$ is the proper time along the hotspot trajectory. Since $\mathcal{L}$ is independent of $t$ and $\phi$, the corresponding Euler–Lagrange equations give the conserved quantities
\begin{equation}
    E=f(r)\frac{\mathrm{d}t}{\mathrm{d}\tau},
    \qquad
    L=r^2\frac{\mathrm{d}\phi}{\mathrm{d}\tau},
\end{equation}
where $E$ and $L$ are the specific energy and specific angular momentum. The timelike normalization $2\mathcal{L}=-1$ then gives the radial equation
\begin{equation}
    \left(\frac{\mathrm{d}r}{\mathrm{d}\tau}\right)^2=E^2-V_{\rm eff}(r),
    \qquad
    V_{\rm eff}(r)=f(r)\left(1+\frac{L^2}{r^2}\right).
\end{equation}
To describe the radial motion, we introduce the Darwin parametrization~\cite{Darwin:1961}
\begin{equation}
r(\chi)=\frac{pM}{1+e\cos\chi},\label{eq:darwin_radius}
\end{equation}
where $p$ is the dimensionless semilatus rectum, $e$ is the orbital eccentricity, and $\chi$ parametrizes the radial phase. The periastron and apastron radii are
\begin{equation}
    r_{\rm p}=\frac{pM}{1+e},
    \qquad
    r_{\rm a}=\frac{pM}{1-e},
\end{equation}
which give $e=(r_{\rm a}-r_{\rm p})/(r_{\rm a}+r_{\rm p})$. We consider bound, noncircular orbits with $0<e<1$ and $p>6+2e$.
At the radial turning points, the condition $E^2=V_{\rm eff}(r_{\rm p})=V_{\rm eff}(r_{\rm a})$ determines the conserved quantities in terms of $p$ and $e$
\begin{equation}
    E^2=\frac{(p-2)^2-4e^2}{p(p-3-e^2)},
    \qquad
    L^2=\frac{M^2p^2}{p-3-e^2}.
\end{equation}
Using $\mathrm{d}r/\mathrm{d}\chi=pMe\sin\chi/(1+e\cos\chi)^2$ in the radial equation, and choosing $\chi$ to increase along the motion, we obtain
\begin{equation}
    \frac{\mathrm{d}\tau}{\mathrm{d}\chi}=\frac{Mp^{3/2}\sqrt{p-3-e^2}}{(1+e\cos\chi)^2\sqrt{p-6-2e\cos\chi}}.
\end{equation}
Combining this expression with the conserved energy and angular momentum gives
\begin{equation}
    \frac{\mathrm{d}\phi}{\mathrm{d}\chi}=\frac{\sqrt{p}}{\sqrt{p-6-2e\cos\chi}},
    \label{eq:darwin_phi}
    \end{equation}
\begin{equation}
    \frac{\mathrm{d}t}{\mathrm{d}\chi}=\frac{Mp^2\sqrt{(p-2)^2-4e^2}}{(p-2-2e\cos\chi)(1+e\cos\chi)^2    \sqrt{p-6-2e\cos\chi}}.
    \label{eq:darwin_time}
\end{equation}
Integrating Eqs.~(\ref{eq:darwin_time}) and (\ref{eq:darwin_phi}) and using Eq.~(\ref{eq:darwin_radius}) gives the hotspot coordinates $(t(\chi),r(\chi),\pi /2,\phi(\chi))$ in the equatorial plane.
The interval $0\leq\chi\leq2\pi$ describes one radial cycle from periastron through apastron to the next periastron. Integrating over this interval gives the radial period $T_r$ and the accumulated azimuthal angle $\Delta\phi_r$
\begin{equation}
\begin{aligned}
    T_r&=\int_0^{2\pi}\frac{\mathrm{d}t}{\mathrm{d}\chi}\,\mathrm{d}\chi,
    \\[4pt]
    \Delta\phi_r&=\int_0^{2\pi}\frac{\sqrt{p}}{\sqrt{p-6-2e\cos\chi}}\,\mathrm{d}\chi.
\end{aligned}
\end{equation}

\begin{figure}[htbp]
    \centering

    \begin{subfigure}[t]{0.3\linewidth}
        \centering
        \caption{$(3,0,1)$}
        \includegraphics[width=\linewidth]{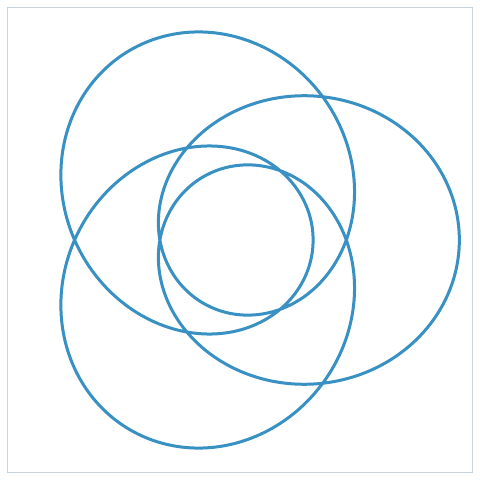}
        \label{fig:30105}
    \end{subfigure}
    \hspace{0.01\linewidth}
    \begin{subfigure}[t]{0.3\linewidth}
        \centering
        \caption{$(3,1,1)$}
        \includegraphics[width=\linewidth]{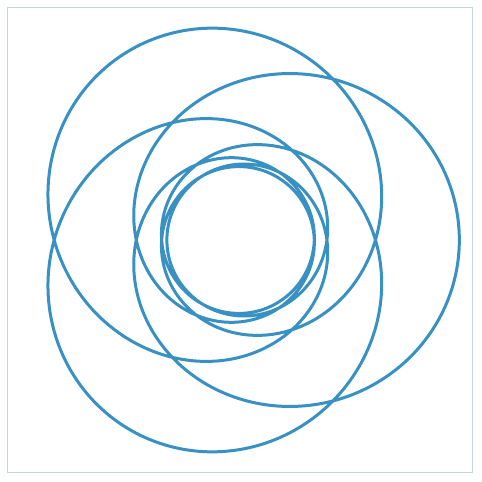}
        \label{fig:31105}
    \end{subfigure}

    \vspace{0.01cm}

    \begin{subfigure}[t]{0.3\linewidth}
        \centering
        \caption{$(4,0,1)$}
        \includegraphics[width=\linewidth]{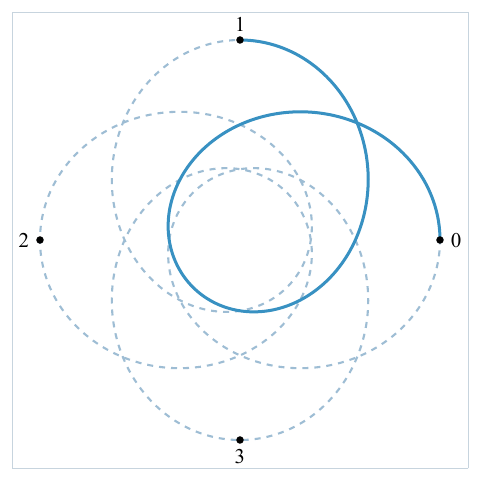}
        \label{fig:40105}
    \end{subfigure}
    \hspace{0.01\linewidth}
    \begin{subfigure}[t]{0.3\linewidth}
        \centering
        \caption{$(4,0,3)$}
        \includegraphics[width=\linewidth]{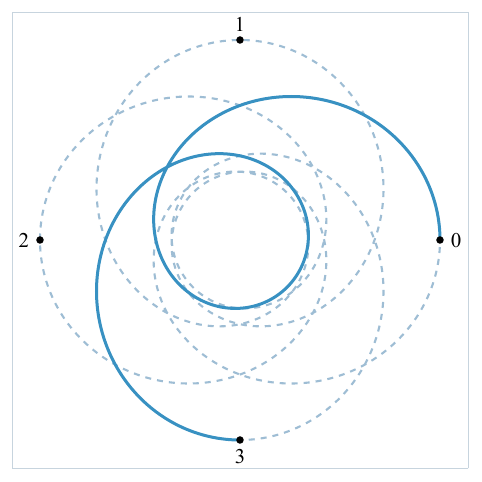}
        \label{fig:40305}
    \end{subfigure}

    \caption{
    Examples of bound closed geodesic orbits in Schwarzschild spacetime classified by the integers $(z,w,v)$, with $e=0.5$. In panels (c) and (d), the numbered points mark the apastron vertices; the solid curve shows one radial period from vertex $0$ to the next apastron, while the remainder of the closed orbit is shown by the dashed curve. Each panel is independently rescaled for clarity.
     }
    \label{fig:closed_orbit_classification}
\end{figure}

In general, $\Delta\phi_r$ is not an integer multiple of $2\pi$, so the hotspot does not return to its starting position after a single radial period. 
A closed spatial trajectory is obtained when the accumulated angle becomes an integer multiple of $2\pi$ after a finite number of radial periods~\cite{Levin:2008mq}.
The resulting closed trajectories can have different numbers of leaves and additional whirls near periastron.
Following Levin and Perez-Giz~\cite{Levin:2008mq}, we describe these structures using three integers $(z,w,v)$, illustrated in Fig.~\ref{fig:closed_orbit_classification}. The integer $z$ counts the leaves of the complete closed orbit, each corresponding to one radial cycle. The integer $w$ counts the additional full whirls around the black hole during each radial period. For example, the $(3,0,1)$ and $(3,1,1)$ orbits in Fig.~\ref{fig:30105} and~\ref{fig:31105} both have three leaves, but the latter executes one additional whirl during each radial cycle.

The leaf and whirl numbers alone do not specify the order in which the leaves are traced. To describe this order, we label the apastron vertices by $0,1,\ldots,z-1$ in the direction of orbital motion. The third integer $v$ gives the label of the next apastron reached after the reference vertex $0$.
For the $(4,0,1)$ orbit in Fig.~\ref{fig:40105}, the vertices are visited in the sequence $0\to1\to2\to3\to0$, whereas the $(4,0,3)$ orbit in Fig.~\ref{fig:40305} follows $0\to3\to2\to1\to0$. For the multileaf orbits considered here, $z>1$, $w\geq0$, and $1\leq v\leq z-1$. To eliminate degeneracy in the orbital labeling, $z$ and $v$ are required to be relatively prime, i.e., $\gcd(z,v)=1$~\cite{Levin:2008mq}.

This geometric classification determines the azimuthal advance during one radial period,
\begin{equation}
\Delta\phi_r=2\pi\left(1+w+\frac{v}{z}\right).
\label{eq:azimuthal_closure}
\end{equation}
The orbit closes after a time interval $T=zT_r$, with a total accumulated azimuthal angle of
\begin{equation}
\Delta\phi_{\rm tot}=z\Delta\phi_r=2\pi\left[z(1+w)+v\right].
\end{equation}
The orbital topology and shape are specified by $(z,w,v)$ and $e$.
We construct the corresponding geodesics numerically subject to the closure condition in Eq.~\eqref{eq:azimuthal_closure}.
These trajectories provide the hotspot motion used to construct the image correlations in the next subsection.

\subsection{Correlations of hotspot images}

Correlation analysis characterizes the relationships between signals at different times and positions and has been used to investigate accretion variability and strong-lensing signatures~\cite{Gandhi:2008qr,Fukumura_2010,Hadar:2020fda,Conroy:2023kec}.
In particular, autocorrelation compares the same signal at different times or positions and can reveal recurring patterns.
For the orbiting hotspots considered here, we retain both the time lag and angular displacement to investigate how their orbital motion is reflected in the image correlations.

The geodesics constructed in the preceding subsection specify the hotspot position $\mathbf{x}_{\rm s}(t_{\rm s}(\chi))=(r(\chi),\pi/2,\phi(\chi))$, where $t_{\rm s}(\chi)$ denotes the emission time. 
For a static observer at $\mathbf{x}_{\rm o}=(r_{\rm o},\theta_{\rm o},\phi_{\rm o})$, we obtain the hotspot images by tracing null geodesics connecting the emission events to the observer, following the methods developed in Refs.~\cite{Zhu:2024vxw,Zhu:2025jqh}.
The ray-tracing procedure maps the hotspot position to the celestial coordinates of its images
\begin{equation}
    \mathrm{RT}:
    (\mathbf{x}_{\rm o},\mathbf{x}_{\rm s})
    \mapsto
    (\Psi,\Phi),
    \label{eq:image_mapping}
\end{equation}
where $(\Psi,\Phi)$ are celestial coordinates on the observer's sky.
The photon arrival time is obtained by adding the light-travel time to the emission time.
Using the coordinate-time relation in Eq.~(5) of Ref.~\cite{Zhu:2025jqh}, we write
\begin{equation}
    t=t_{\rm s}\pm_r\int_{r_{\rm s}}^{r_{\rm o}}\frac{r\,\mathrm{d}r}{f(r)\sqrt{r^2-\rho^2 f(r)}} ,
    \label{eq:photon_arrival_time}
\end{equation}
where $r_{\rm s}=r(\chi)$ is the radius at the emission time $t_{\rm s}(\chi)$ and $\rho$ is the photon impact parameter, related to the celestial coordinate by $\rho=r_{\rm o}\sin\Psi/\sqrt{f(r_{\rm o})}$.
The integral is evaluated piecewise along the photon path, with the positive sign for outward segments and the negative sign for inward segments, splitting the integral at any radial turning point.
For fixed observer position $\textbf{x}_\text{o}$ and the given hotspot trajectory $\textbf{x}_s(t_s)$, the apparent track of each image is obtained as $(\Psi_p (t),\Phi_p (t))$ via Eqs.~(\ref{eq:image_mapping}) and (\ref{eq:photon_arrival_time}). The following calculations concern the primary image.

The observed specific intensity at frequency $\nu$ is given by \cite{Zhu:2025jqh}
\begin{equation}
    I_{\rm obs} (\Psi,\Phi;t,\mathbf{x}_{\rm o})=g^3(\Psi,\Phi;\mathbf{x}_{\rm s},\mathbf{x}_{\rm o}) I_{\rm emt}(t_{\rm s},\mathbf{x}_{\rm s}),
    \label{eq:redshift_intensity}
\end{equation}
where $g$ is the ratio of observed to emitted frequency, and $I_{\rm emt}$ is the emission intensity of a hotspot.
Integrating the observed intensity over solid angle gives the specific flux of the primary image,
\begin{equation}
    F_\nu(t)=\int I_{\rm obs}(\Psi,\Phi;t,\mathbf{x}_{\rm o})\,\mathrm{d}\Omega,
\label{eq:image_flux}
\end{equation}
where $\mathrm{d}\Omega = \sin\Psi\,\mathrm{d}\Psi\,\mathrm{d}\Phi$. Lensing effects enter through the mapping of the source onto the observer's sky.
For the point-like hotspot considered here, the image position and flux determine the observed intensity distribution through Eq.~(13) of Ref.~\cite{Zhu:2025jqh},
\begin{equation}
    I_\text{obs} \left( \Psi, \Phi ; t , \textbf{x}_{\text{o}} \right) = \frac{F_\nu (t)}{\sin\Psi} \delta\left(\Psi -\Psi_p (t)\right)\delta\left(\Phi - \Phi_p(t)\right)~, \label{5}
\end{equation}
where $\delta$ denotes the Dirac delta function, with the azimuthal argument understood modulo $2\pi$.
Substituting the hotspot intensity from Eq.~(\ref{5}), the intensity correlation takes the form \cite{Zhu:2025jqh}
\begin{eqnarray}
    C(\Delta t,\Delta\Phi) &=& \left\langle I_{\text{obs}}  \left( \Psi, \Phi ; t , \textbf{x}_{\text{o}} \right) I_{\text{obs}}  \left( \bar{\Psi}, \Phi + \Delta \Phi ; t + \Delta t, \textbf{x}_{\text{o}} \right) \right\rangle \nonumber \\ &=&\sum_{t_*}\frac{F_{\nu} (t_*)F_{\nu} (t_*+\Delta t)}{\left| \dot{\Phi}_{p} (t_*) - \dot{\Phi}_{p} (t_*+\Delta t) \right|}~,
    \label{eq:point_source_correlation}
\end{eqnarray}
where the brackets denote integration over the reference arrival time  and celestial coordinates, and an overdot denotes differentiation with respect to $t$.
Here, $\Delta t$ and $\Delta\Phi$ denote the arrival-time difference and apparent azimuthal displacement, respectively.
The sum runs over all arrival times $t_*$ within the sampled reference interval that satisfy the angular matching condition \cite{Zhu:2025jqh}.
Accounting explicitly for the $2\pi$ periodicity of the azimuthal coordinate, this condition reads:
\begin{equation}
    \Delta\Phi+\Phi_p (t_*)-\Phi_p (t_*+\Delta t)=2\pi k,
    \qquad k\in\mathbb{Z}.
    \label{matching_condition}
\end{equation}
At a fixed time lag $\Delta t$, different orbital phases can contribute at different angular displacements $\Delta\Phi$. The resulting correlation pattern in the $(\Delta t,\Delta\Phi)$ plane therefore characterizes the temporal and angular structure of the image motion.
Our main analysis focuses on the autocorrelation of the primary images of the hotspot. Correlations involving higher-order images are presented and discussed in Appendix~\ref{A}.

In our numerical calculations, the hotspot is initialized at periastron, and the reference arrival times are sampled over one complete orbital period $T$. Owing to the $2\pi$ periodicity of the azimuthal coordinate, the angular displacement is mapped onto the principal interval $\Delta\Phi\in[-\pi,\pi)$. For the primary-image autocorrelations presented in the main text, each correlation is normalized by its own maximum and displayed with a power-law transformation,
\begin{equation}
    \overline{C} (\Delta t,\Delta\Phi)\equiv\left[\frac{C (\Delta t,\Delta\Phi)}{  C_{\max} }\right]^{\gamma},
    \qquad \gamma=0.1,
\end{equation}
where $C_{\max} $ is the maximum value over the displayed correlation. This independent normalization facilitates comparison of the correlation morphology but does not preserve relative correlation amplitudes across maps. The power-law transformation makes weaker features more visible. The resulting patterns form the basis of the analysis of orbital topology in the next section.

\section{Correspondence Between Closed-Orbit Topology and Correlation Morphology}\label{sec:topology_correlations}

The motion along the closed orbits described above produces time-dependent images on the observer's sky, which might encode information about the underlying spacetime geometry.
In this section, we calculate the correlations associated with closed hotspot orbits, characterize their correlation-band structures, and examine how they encode the orbital topology $(z,w,v)$. 
Specifically, we focus on the autocorrelation of the primary image and analyze the number and arrangement of correlation bands as functions of the orbital topology $(z,w,v)$.

To explore correlation signatures of the orbital topology $(z,w,v)$, we fix the orbital eccentricity at $e=0.3$ and the observer inclination at $i=\pi/9$ in this section.
We define $N_{\rm band}$ as the number of correlation bands within one orbital period after excluding the two boundary bands passing through $ (\Delta t,\Delta\Phi)=(0,0) $ and $(T,0)$. This counting convention is used throughout the following analysis.
By calculating the primary-image autocorrelation $\overline{C} $ for different combinations of $(z,w,v)$, we find the empirical relation as follows,
\begin{equation}
    N_{\rm band}=z(w+1)+v-1={}\frac{\Delta\phi_{\rm tot}}{2\pi}-1~,
    \label{eq:empirical_band_number}
\end{equation}
where $\Delta\phi_{\rm tot}$ is the total accumulated azimuthal angle over one complete closed orbit. We illustrate this relation below by varying one topological parameter at a time.

\begin{figure}
    \centering
    \includegraphics[width=1\linewidth]{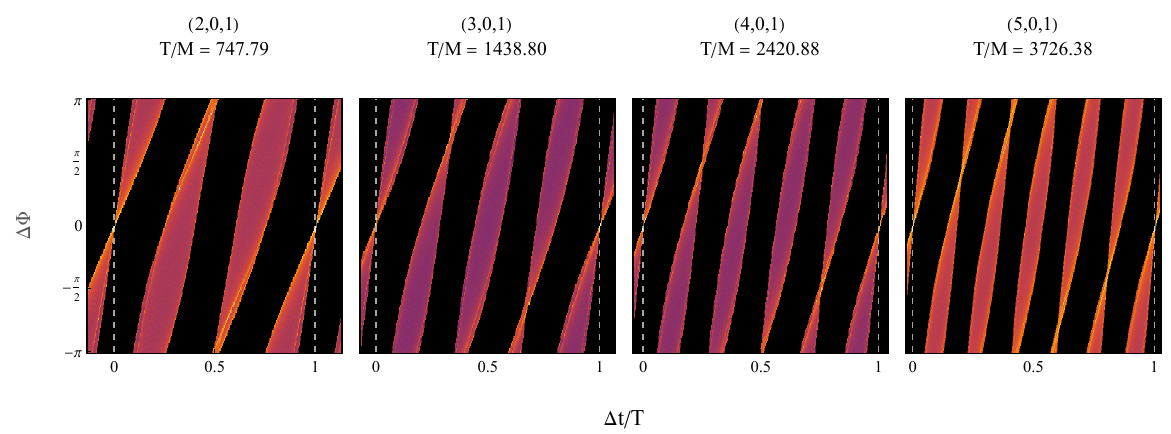}
    \caption{Primary-image autocorrelations for closed hotspot orbits with different leaf numbers $z$. From left to right, the four panels correspond to $(2,0,1)$, $(3,0,1)$, $(4,0,1)$, and $(5,0,1)$, respectively. The two vertical dashed lines in each panel delimit one full orbital period $T$, with the corresponding value of $T/M$ indicated above the panel. The horizontal axis shows the normalized time lag $\Delta t/T$.
}
    \label{diff_z}
\end{figure}

We first fix $w=0$ and $v=1$ and vary the number of leaves $z$ from $2$ to $5$ in Fig.~\ref{diff_z}. 
Within one orbital period, these orbits contain $2$, $3$, $4$, and $5$ correlation bands, respectively, according to the counting convention above.
Thus, for the $(z,0,1)$ family examined here, we have $N_{\rm band}=z$.
\begin{figure}
    \centering
    \includegraphics[width=1\linewidth]{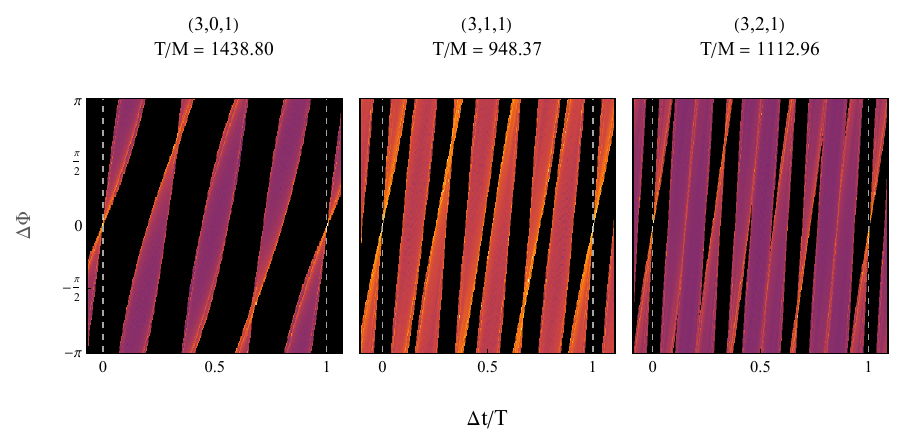}
    \caption{Primary-image autocorrelations for closed hotspot orbits with different whirl numbers $w$. From left to right, the three panels correspond to $(3,0,1)$, $(3,1,1)$, and $(3,2,1)$, respectively. The two vertical dashed lines in each panel delimit one full orbital period $T$, with the corresponding value of $T/M$ indicated above the panel.}
    \label{diff_w}
\end{figure}
We next fix $z=3$ and $v=1$ and vary the whirl number $w$. As shown in Fig.~\ref{diff_w}, the $(3,0,1)$, $(3,1,1)$, and $(3,2,1)$ orbits produce $3$, $6$, and $9$ correlation bands, respectively. The band number increases by $z$ for each additional whirl. The sequence satisfies the relation $N_{\rm band}=z(w+1)$.
\begin{figure}
    \centering
    \includegraphics[width=1\linewidth]{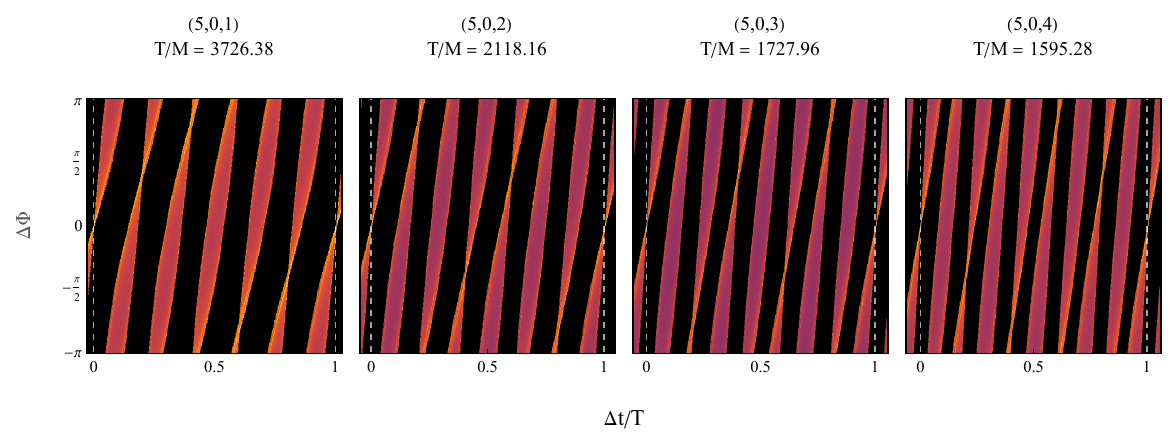}
    \caption{Primary-image autocorrelations for closed hotspot orbits with different vertex parameters $v$. From left to right, the four panels correspond to $(5,0,1)$, $(5,0,2)$, $(5,0,3)$, and $(5,0,4)$, respectively. The two vertical dashed lines in each panel delimit one full orbital period $T$, with the corresponding value of $T/M$ indicated above the panel.}
    \label{diff_v}
\end{figure}
Finally, we examine the dependence on the vertex parameter $v$ while fixing $z=5$ and $w=0$. Although these orbits have the same number of leaves, they visit the apastron vertices in different orders. Fig.~\ref{diff_v} shows that the $(5,0,1)$, $(5,0,2)$, $(5,0,3)$, and $(5,0,4)$ orbits produce $5$, $6$, $7$, and $8$ correlation bands, respectively. This sequence is consistent with the relation $N_{\rm band}=z+v-1$.
To test the empirical relation in Eq.~\eqref{eq:empirical_band_number} beyond the one-parameter orbit sequences, we additionally examined the $(3,1,2)$, $(3,2,2)$, $(4,1,3)$, and $(5,1,2)$ orbits, shown in Appendix~\ref{app:additional_checks}. Their band numbers are $7$, $10$, $10$, and $11$, respectively, all of which agree with Eq.~\eqref{eq:empirical_band_number}.

The band count $N_\text{band}$ alone, however, does not uniquely determine the orbital topology. For example, 
the $(3,0,2)$ and $(4,0,1)$ orbits both contain four correlation bands under our counting convention.
This is a degeneracy in the band count and does not imply that the two correlations are otherwise identical. It motivates us to identify an additional feature that can distinguish between these two cases.

\begin{figure}[htbp]
    \centering

    \begin{subfigure}[t]{0.47\linewidth}
        \centering
        \caption{$(3,0,2)$}
        \includegraphics[width=\linewidth]{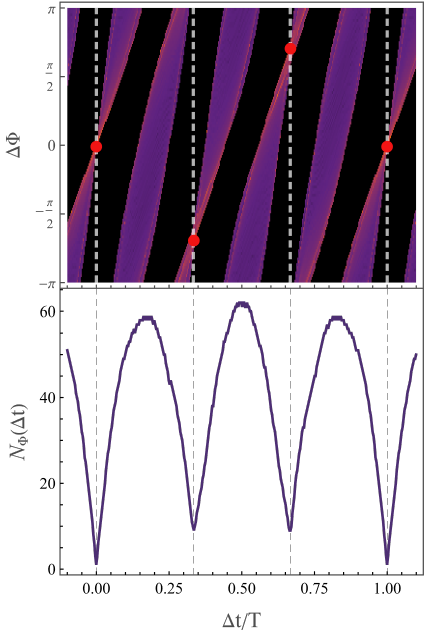}
        \label{fig:corr302}
    \end{subfigure}
    \hfill
    \begin{subfigure}[t]{0.47\linewidth}
        \centering
        \caption{$(4,0,1)$}
        \includegraphics[width=\linewidth]{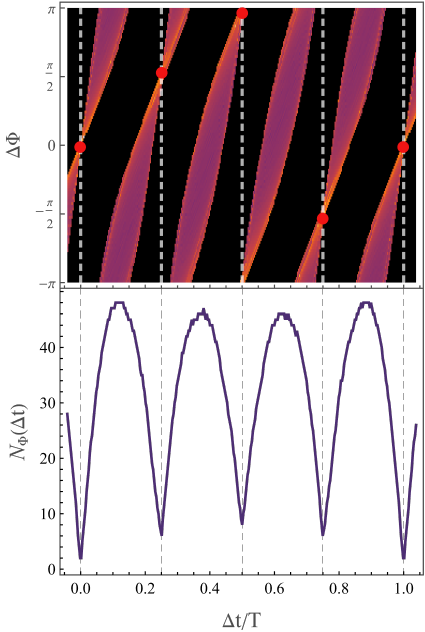}
        \label{fig:corr401}
    \end{subfigure}
    \caption{Primary-image autocorrelations and the corresponding angular-bin counts for the $(3,0,2)$ and $(4,0,1)$ closed hotspot orbits. In each panel, the upper plot shows the autocorrelation as a function of the normalized time lag $\Delta t/T$ and angular displacement $\Delta\Phi$, while the lower plot shows $N_{\Phi}(\Delta t)$. The red dots mark the recurrence points in the autocorrelations, and the vertical dashed lines indicate their approximate locations, $\Delta t_k/T=k/z$ for $k=0,\ldots,z$. These recurrence points correspond to local minima of $N_{\Phi}(\Delta t)$. The eccentricity is fixed at $e=0.3$, and the hotspot is initialized at periastron.}
    \label{fig:band_recurrence}
\end{figure}

We find that the correlation bands repeatedly converge toward localized points in the $(\Delta t,\Delta\Phi)$ plane, which we refer to as recurrence points. To identify these points quantitatively, for each time lag $\Delta t$ we count the number of angular bins containing a nonzero correlation signal and denote it by $N_{\Phi}(\Delta t)$, as shown in the lower panels of Fig.~\ref{fig:band_recurrence}. The recurrence points are identified by local minima of $N_{\Phi}(\Delta t)$, where the correlation signal occupies a smaller angular range.
A closed orbit with $z$ leaves completes $z$ radial cycles during one full orbital period $T$. Its radial period is therefore $T_r=T/z$. In our numerical results, the recurrence points occur near the time lags
\begin{equation}
    \Delta t_k^{\rm rec} = \frac{k}{z}T,
    \qquad k=0,1,\ldots,z,
\end{equation}
Their approximate temporal spacing therefore corresponds to the radial period of the hotspot motion.
On the closed interval $0\leq\Delta t\leq T$, the number of displayed recurrence points is 
\begin{equation}
    N_{\rm rec}=z+1 , \label{eq:rec}
\end{equation}
where both endpoints are included.

This recurrence structure breaks the band-count degeneracy. 
As shown in the upper panels of Fig.~\ref{fig:band_recurrence}, the $(3,0,2)$ and $(4,0,1)$ orbits have the same number of correlation bands, $N_\text{band}^{(3,0,2)}=N_\text{band}^{(4,0,1)}=4$, but exhibit different numbers of recurrence points, with $N_\text{rec}^{(3,0,2)}=4$ and $N_\text{rec}^{(4,0,1)}=5$, respectively. 
For the orbit families examined, varying either $w$ or $v$ at fixed $z$ changes the number and arrangement of the correlation bands but does not change the number of recurrence points. The angular locations of the recurrence points may also shift, while their temporal spacing remains approximately $T/z$. Additional checks for the $(3,2,2)$ and $(4,1,3)$ orbits yield $4$ and $5$ displayed recurrence points, shown in Appendix~\ref{app:additional_checks}, respectively. These results remain consistent with $N_{\rm rec}=z+1$ when both $w$ and $v$ are nontrivial.

Combining the empirical counting relations in Eqs.~(\ref{eq:empirical_band_number}) and (\ref{eq:rec}) with the constraints of the Levin–Perez-Giz orbit taxonomy \cite{Levin:2008mq}, namely, $w\geq0$, $1\leq v\leq z-1$, and $\gcd(z,v)=1$, the orbital topology can be extracted from the correlation structures, namely, 
\begin{equation}
\begin{aligned}
    z &= N_{\rm rec}-1, \\[3pt]
    w &= \left\lfloor
         \frac{N_{\rm band}+1}{N_\text{rec}-1}
         \right\rfloor-1, \\[3pt]
    v &= (N_{\rm band}+1)\bmod (N_\text{rec}-1),
\end{aligned}
\label{eq:topology_from_counts}
\end{equation}
where $\lfloor x\rfloor$ denotes the floor function, and $\bmod$ denotes the remainder operation. 
The band number depends on the combined effects of $z$, $w$, and $v$ (Eq.~(\ref{eq:empirical_band_number})), whereas the recurrence count determines the leaf number $z$ (Eq.~(\ref{eq:rec})). 
These two features recover the orbital topology for the orbit families examined here. 
Having established this correspondence at fixed eccentricity, we next examine how the correlation structures change when the eccentricity is varied.

\section{Effects of orbital eccentricity in correlations}\label{sec:effect_ecc}

In addition to the discrete topological parameters $(z,w,v)$, the geometry and dynamical properties of a closed orbit depend continuously on the orbital eccentricity $e$. To isolate the effect of eccentricity, we fix the orbital topology at $(z,w,v)=(3,0,1)$ and vary $e$. We retain the observer inclination $i=\pi/9$, and the hotspot is initialized at periastron in all cases.
\begin{figure}
    \centering
    \includegraphics[width=0.8\linewidth]{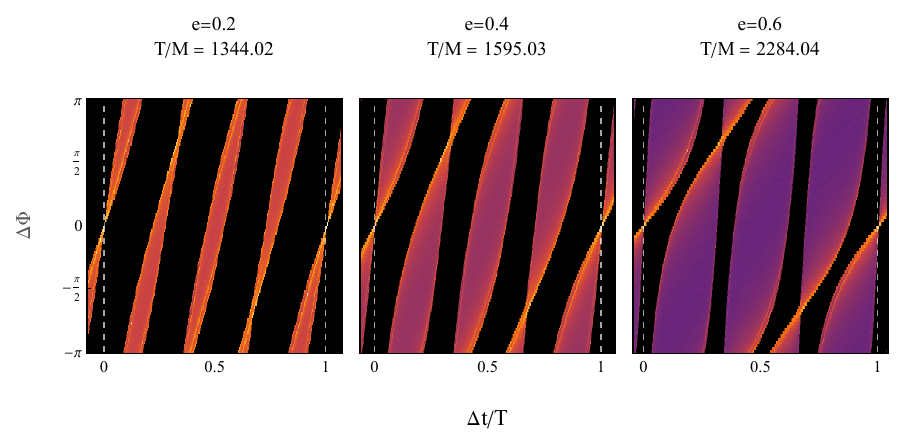}
    \caption{Primary-image autocorrelations for closed hotspot orbits with different orbital eccentricities $e$, with the orbital topology fixed at $(z,w,v)=(3,0,1)$. From left to right, the three panels correspond to $e=0.2$, $e=0.4$, and $e=0.6$, respectively. The corresponding value of $T/M$ is indicated above each panel.}
    \label{diff_e}
\end{figure}
As shown in Fig.~\ref{diff_e}, when the time lag is normalized by the orbital period, the correlation bands occupy a larger temporal fraction as the eccentricity increases. 
A larger eccentricity produces a greater radial excursion and a more nonuniform orbital motion.
For a given angular displacement, the normalized time lag depends on orbital phase. Greater variation at higher eccentricity may explain the broader bands.

To quantify this effect, we measure the temporal width of the bands along the angular slice $\Delta\Phi=0$. In the numerical correlations, this slice is represented by the angular grid point closest to zero. 
Following the counting convention introduced in Sec.~\ref{sec:topology_correlations}, we exclude the two boundary bands passing through $(\Delta t,\Delta\Phi)=(0,0)$ and $(T,0)$ and measure the widths of the remaining three bands.
The mean temporal width is defined as
\begin{equation}
    \overline{W}_{t}=\frac{1}{N_{\rm band}}\sum_{k=1}^{N_{\rm band}} W_{t,k},
    \label{eq:mean_temporal_width}
\end{equation}
where $W_{t,k}$ is the temporal width of the $k$th contiguous interval with nonzero correlation on the selected slice, and $N_{\rm band}=3$.  
Since the orbital period varies significantly with eccentricity, we further define the normalized mean temporal width
\begin{equation}
    \mathcal{W}_{t}\equiv\frac{\overline{W}_{t}}{T},
    \label{eq:normalized_temporal_width}
\end{equation}
which allows a direct comparison among orbits with different periods.
\begin{figure}
    \centering
    \includegraphics[width=0.7\linewidth]{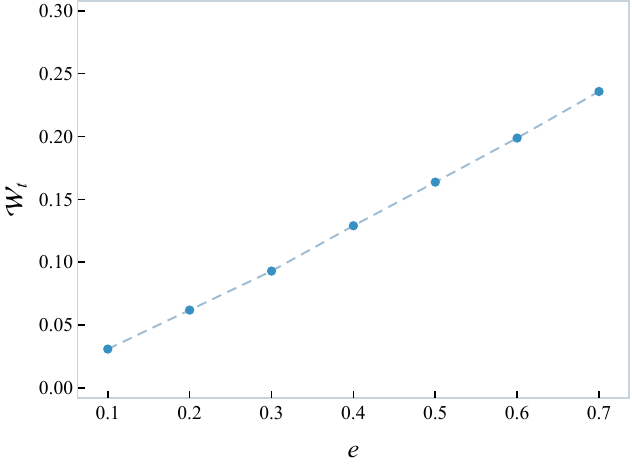}
    \caption{Normalized mean temporal width $\mathcal{W}_{t}$ of the correlation bands as a function of the orbital eccentricity $e$ for the $(3,0,1)$ closed hotspot orbit. The dashed line connects adjacent data points and is included only to guide the eye.
}
    \label{fig:width_e}
\end{figure}
The resulting normalized widths are presented in Fig.~\ref{fig:width_e}. For the eccentricities shown in Fig.~\ref{diff_e}, the band count and qualitative recurrence structure are retained as the bands broaden. Thus, for the $(3,0,1)$ orbit family considered here, eccentricity changes the temporal extent of the correlation bands while preserving the characteristic structure used to identify the orbital topology.

\section{Conclusions and discussions}\label{sec:conclusions}

In this work, we investigated how the topology of bound closed hotspot orbits around a Schwarzschild black hole is reflected in the spatiotemporal autocorrelations of the primary image.
For all orbit families examined here, the number of correlation bands within one orbital period is consistent with the empirical relation $N_{\rm band}=z(w+1)+v-1$.
The autocorrelations also exhibit recurrence points at which the correlation bands converge. Over the interval $0\leq\Delta t\leq T$, the number of recurrence points is $N_{\rm rec}=z+1$ when both endpoints are included.
The band count depends on all three integers $(z,w,v)$, whereas the recurrence count depends only on $z$. These two counts $(N_\text{band},N_\text{rec})$ allow the orbital topology to be recovered using Eq.~(\ref{eq:topology_from_counts}), providing a quantitative method for recovering the multileaf orbital structure from hotspot-image autocorrelations. 
We also examined the effect of eccentricity while fixing the topology at $(3,0,1)$. 
Increasing the eccentricity broadens the correlation bands in normalized time lag while preserving the band count and qualitative recurrence structure for the eccentricities examined. 

The present findings are based on numerical calculations for a pointlike hotspot on equatorial closed geodesics in Schwarzschild spacetime, at a fixed observer inclination. 
A natural extension is to examine whether closed orbits with the same $(z,w,v)$ in different spacetimes produce distinguishable correlations.
Extending the correlation analysis to equatorial closed orbits in Kerr spacetime and selected modified-gravity models would test whether the same counting relations hold.
Even if the counts remain unchanged, differences in orbital dynamics and light propagation may appear in the widths and shapes of the correlation bands. Whether these features can distinguish the underlying spacetime from Schwarzschild, after accounting for the effects of eccentricity and observer inclination, remains to be investigated.

\smallskip
{\it Acknowledgments}:  This work has been supported by the National Natural
Science Foundation of China Grants (No. 12305073, No. 12347101, and No. 12275034).

\appendix

\section{Auto- and cross-correlations for higher-order images\label{A}}
In this study, we focus on the autocorrelation of the primary images of the orbital hotspot. However, it is well known that photons emitted from the hotspot can wind around the black hole multiple times before reaching the observer, referred to as multiple images \cite{Luminet:1979}. These images are classified by the image order $n$, which corresponds to the number of half-orbits of the light path connecting the source and the observer. Previous studies have demonstrated that the correlations from secondary and higher-order images are also informative and can be used to extract black hole parameters \cite{Zhang:2025vyx,Zhu:2025jqh}. Here, we will show that the higher-order correlations are not essential, as the primary-image correlation alone is sufficient to fully characterize the orbital topology.

\begin{figure}[H]
    \centering
    \includegraphics[width=0.7\linewidth]{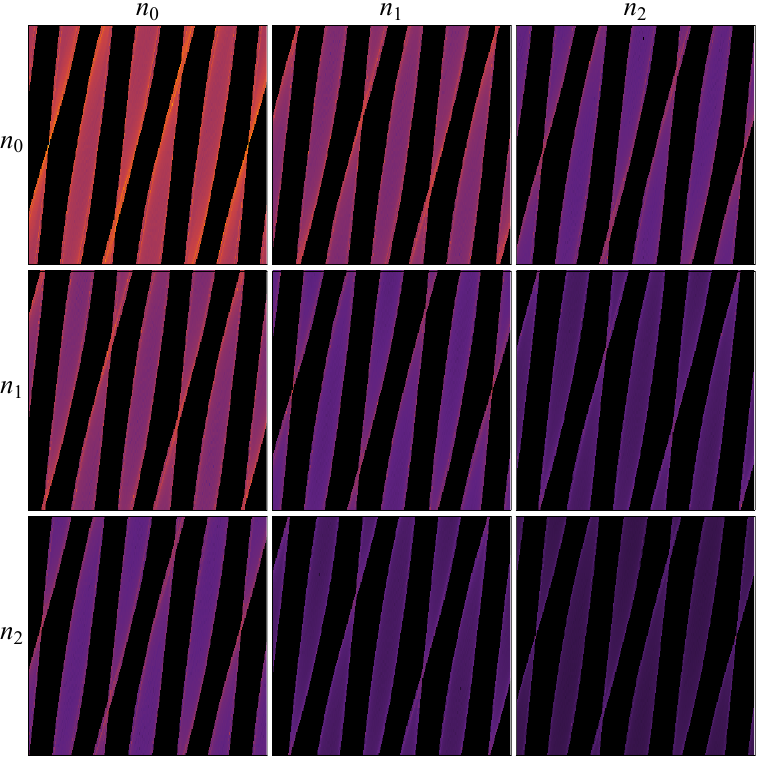}
    \caption{Dimensionless auto- and cross-correlations, $\overline{C}^{(n_1,n_2)}(\Delta t,\Delta\Phi)$, for a closed hotspot orbit with topology $(z,w,v)=(3,0,2)$. Here, $n_1, n_2=0$ denotes the primary image and $n_1,n_2\geq1$ denotes higher-order images. When $n_1=n_2$, it gives the autocorrelation of a single image order; when $n_1\neq n_2$, it gives the cross-correlation between different image orders.  The rows and columns specify the first and second image orders, respectively, each ranging from 0 to 2.  All panels are normalized by the maximum of the primary-image autocorrelation $C $ and displayed using the exponent $\gamma=0.07$.}
    \label{fig:302auto-corss}
\end{figure}

Figure~\ref{fig:302auto-corss} presents auto- and cross-correlations for image orders $n_1,n_2=0,1,2$, using the $(3,0,2)$ orbit as a representative example.
In this example, the characteristic multiband structure remains qualitatively similar across the image pairs shown, although the detailed band shapes and angular locations vary.

\section{Additional checks of the counting relations}\label{app:additional_checks}

\begin{figure}
    \centering
    \includegraphics[width=1\linewidth]{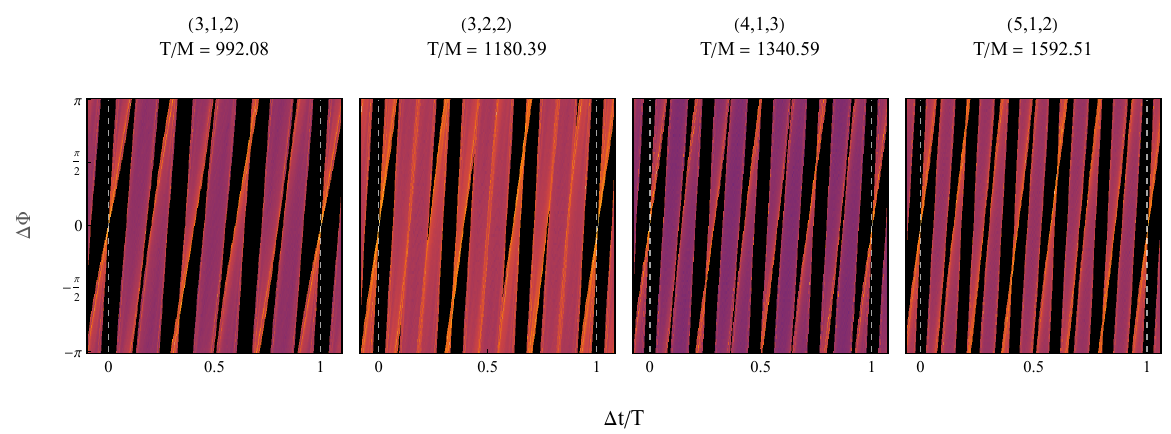}
    \caption{Primary-image autocorrelations for closed hotspot orbits with different combinations of the parameters $(z,w,v)$, at $e=0.3$. From left to right, the four panels correspond to $(3,1,2)$, $(3,2,2)$, $(4,1,3)$, and $(5,1,2)$, respectively. The plotting conventions are the same as in Fig.~\ref{diff_z}.}
    \label{fig:additional_bands}
\end{figure}

We present additional numerical checks of the counting relations discussed in Sec.~\ref{sec:topology_correlations}. Fig.~\ref{fig:additional_bands} shows the primary-image autocorrelations for the $(3,1,2)$, $(3,2,2)$, $(4,1,3)$, and $(5,1,2)$ orbits. Following the counting convention used in the main text, we exclude the two boundary bands passing through $(\Delta t,\Delta\Phi)=(0,0)$ and $(T,0)$. The remaining band counts are $7$, $10$, $10$, and $11$, respectively, in agreement with Eq.~\eqref{eq:empirical_band_number}.

\begin{figure}[htbp]
    \centering

    \begin{subfigure}[t]{0.47\linewidth}
        \centering
        \caption{$(3,2,2)$}
        \includegraphics[width=\linewidth]{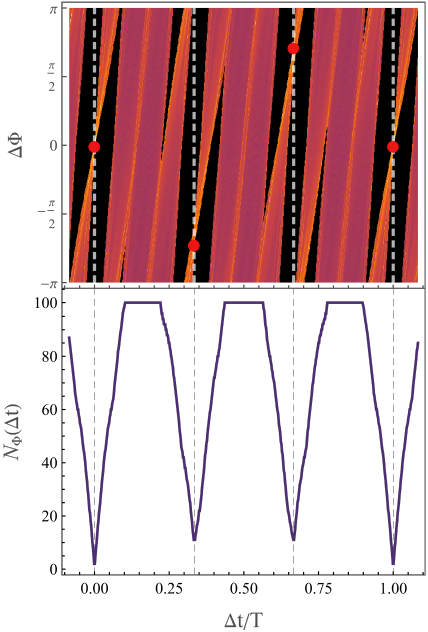}
        \label{fig:322}
    \end{subfigure}
    \hfill
    \begin{subfigure}[t]{0.47\linewidth}
        \centering
        \caption{$(4,1,3)$}
        \includegraphics[width=\linewidth]{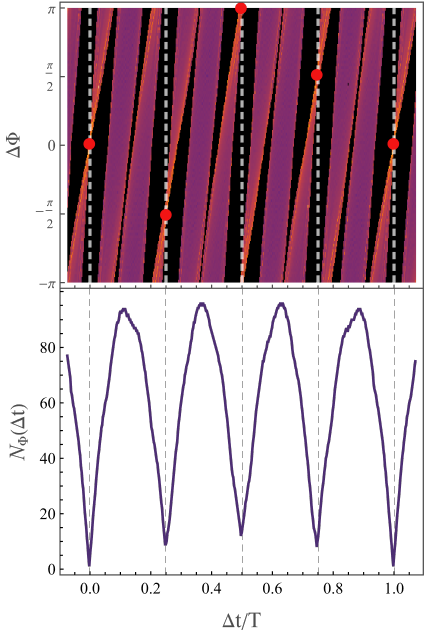}
        \label{fig:413}
    \end{subfigure}
    \caption{Primary-image autocorrelations and the corresponding angular-bin counts for the $(3,2,2)$ and $(4,1,3)$ closed hotspot orbits. The plotting conventions and remaining parameters are the same as in Fig.~\ref{fig:band_recurrence}. Including both endpoints of the interval $[0,T]$, the two orbits exhibit $4$ and $5$ recurrence points, respectively.}
    \label{fig:additional_recurrence}
\end{figure}

We further examine the recurrence structure of the $(3,2,2)$ and $(4,1,3)$ orbits using the angular-bin counts $N_{\Phi}(\Delta t)$ shown in Fig.~\ref{fig:additional_recurrence}. Including both endpoints of $0\leq\Delta t\leq T$, these orbits exhibit $4$ and $5$ recurrence points, respectively, consistent with $N_{\rm rec}=z+1$. Although both orbits have $N_{\rm band}=10$, their different recurrence counts distinguish the two orbital topologies.

\bibliography{ref}

@article{Huang:2026oga,
    author = "Huang, Yi-Han and Guo, Sen and Liang, Yu and Wen, Lin and Lin, Kai",
    title = "{Gravitational waveforms from periodic orbits around Gauss-Bonnet black holes}",
    eprint = "2606.16280",
    archivePrefix = "arXiv",
    primaryClass = "gr-qc",
    month = "6",
    year = "2026"
}

@article{Chen:2022kzv,
    author = "Chen, Yifan and Xue, Xiao and Brito, Richard and Cardoso, Vitor",
    title = "{Photon Ring Astrometry for Superradiant Clouds}",
    eprint = "2211.03794",
    archivePrefix = "arXiv",
    primaryClass = "gr-qc",
    reportNumber = "DESY 22-170",
    doi = "10.1103/PhysRevLett.130.111401",
    journal = "Phys. Rev. Lett.",
    volume = "130",
    number = "11",
    pages = "111401",
    year = "2023"
}

@article{Cardenas-Avendano:2024sgy,
    author = "C{\'a}rdenas-Avenda{\~n}o, Alejandro and Gammie, Charles and Lupsasca, Alexandru",
    title = "{Explanation for the Absence of Secondary Peaks in Black Hole Light Curve Autocorrelations}",
    eprint = "2406.04176",
    archivePrefix = "arXiv",
    primaryClass = "astro-ph.HE",
    doi = "10.1103/PhysRevLett.133.131402",
    journal = "Phys. Rev. Lett.",
    volume = "133",
    number = "13",
    pages = "131402",
    year = "2024"
}

@article{Hadar:2023kau,
    author = "Hadar, Shahar and Harikesh, Sreehari and Chelouche, Doron",
    title = "{Extreme lensing induces spectrotemporal correlations in black-hole signals}",
    eprint = "2305.11247",
    archivePrefix = "arXiv",
    primaryClass = "gr-qc",
    doi = "10.1103/PhysRevD.107.124057",
    journal = "Phys. Rev. D",
    volume = "107",
    number = "12",
    pages = "124057",
    year = "2023"
}

@article{Emami:2023sus,
    author = "Emami, Razieh and others",
    title = "{The EB Correlation in Resolved Polarized Images: Connections to the Astrophysics of Black Holes}",
    eprint = "2305.00387",
    archivePrefix = "arXiv",
    primaryClass = "astro-ph.GA",
    doi = "10.3847/1538-4357/acdc96",
    journal = "Astrophys. J.",
    volume = "955",
    number = "1",
    pages = "6",
    year = "2023"
}

@article{Harikesh:2025nmt,
    author = "Harikesh, Sreehari and Hadar, Shahar and Chelouche, Doron",
    title = "{Exploring lensing signatures through spectrotemporal correlations: Implications for black hole parameter estimation}",
    eprint = "2502.12053",
    archivePrefix = "arXiv",
    primaryClass = "astro-ph.HE",
    doi = "10.1103/wjpm-9byt",
    journal = "Phys. Rev. D",
    volume = "112",
    number = "4",
    pages = "043020",
    year = "2025"
}

@article{Conroy:2025nev,
    author = {Conroy, Nicholas S. and Baub{\"o}ck, Michi and Dhruv, Vedant and Lee, Daeyoung and Chan, Chi-kwan and Joshi, Abhishek V. and Prather, Cora and Gammie, Charles F.},
    title = "{Event Horizon Telescope Pattern Speeds in the Visibility Domain}",
    eprint = "2510.08848",
    archivePrefix = "arXiv",
    primaryClass = "astro-ph.HE",
    doi = "10.3847/1538-4357/ae7326",
    journal = "Astrophys. J.",
    volume = "1005",
    number = "1",
    pages = "71",
    year = "2026"
}

@article{Levin:2008mq,
    author = "Levin, Janna and Perez-Giz, Gabe",
    title = "{A Periodic Table for Black Hole Orbits}",
    eprint = "0802.0459",
    archivePrefix = "arXiv",
    primaryClass = "gr-qc",
    doi = "10.1103/PhysRevD.77.103005",
    journal = "Phys. Rev. D",
    volume = "77",
    pages = "103005",
    year = "2008"
}

@article{Zhu:2024vxw,
    author = "Zhu, Qing-Hua",
    title = "{Observational signatures from higher-order images of moving hotspots in accretion disks}",
    eprint = "2411.04001",
    archivePrefix = "arXiv",
    primaryClass = "gr-qc",
    doi = "10.1103/PhysRevD.111.044010",
    journal = "Phys. Rev. D",
    volume = "111",
    number = "4",
    pages = "044010",
    year = "2025"
}

@article{Darwin:1961,
  author  = {Darwin, C. G.},
  title   = {The Gravity Field of a Particle. II},
  journal = {Proc. R. Soc. Lond. A},
  volume  = {263},
  pages   = {39--50},
  year    = {1961},
  doi     = {10.1098/rspa.1961.0142}
}

@article{Zhu:2025jqh,
    author = "Zhu, Qing-Hua",
    title = "{Auto- and cross-correlations for multiple images of corotating hotspots in accretion disks}",
    eprint = "2503.22343",
    archivePrefix = "arXiv",
    primaryClass = "astro-ph.HE",
    doi = "10.1103/gddn-hzpv",
    journal = "Phys. Rev. D",
    volume = "112",
    number = "6",
    pages = "064021",
    year = "2025"
}

@article{Abuter:2018uum,
    author = "Abuter, R. and others",
    title = "{Detection of orbital motions near the last stable circular orbit of the massive black hole SgrA*}",
    eprint = "1810.12641",
    archivePrefix = "arXiv",
    primaryClass = "astro-ph.GA",
    doi = "10.1051/0004-6361/201834294",
    journal = "Astron. Astrophys.",
    volume = "618",
    pages = "L10",
    year = "2018"
}

@article{GRAVITY:2023data,
    author = "Abuter, R. and others",
    collaboration = "GRAVITY",
    title = "{Polarimetry and astrometry of NIR flares as event horizon scale, dynamical probes for the mass of Sgr A*}",
    eprint = "2307.11821",
    archivePrefix = "arXiv",
    primaryClass = "astro-ph.GA",
    doi = "10.1051/0004-6361/202347416",
    journal = "Astron. Astrophys.",
    volume = "677",
    pages = "L10",
    year = "2023"
}

@article{Zhu:2023omf,
    author = "Zhu, Qing-Hua",
    title = "{Photon ring autocorrelations from gravitational fluctuations around a black hole}",
    eprint = "2301.00913",
    archivePrefix = "arXiv",
    primaryClass = "gr-qc",
    doi = "10.1103/PhysRevD.109.064031",
    journal = "Phys. Rev. D",
    volume = "109",
    number = "6",
    pages = "064031",
    year = "2024"
}

@article{Chesler:2020gtw,
    author = "Chesler, Paul M. and Blackburn, Lindy and Doeleman, Sheperd S. and Johnson, Michael D. and Moran, James M. and Narayan, Ramesh and Wielgus, Maciek",
    title = "{Light echos and coherent autocorrelations in a black hole spacetime}",
    eprint = "2012.11778",
    archivePrefix = "arXiv",
    primaryClass = "gr-qc",
    doi = "10.1088/1361-6382/abeae4",
    journal = "Class. Quant. Grav.",
    volume = "38",
    number = "12",
    pages = "125006",
    year = "2021"
}

@article{Hadar:2020fda,
    author = "Hadar, Shahar and Johnson, Michael D. and Lupsasca, Alexandru and Wong, George N.",
    title = "{Photon Ring Autocorrelations}",
    eprint = "2010.03683",
    archivePrefix = "arXiv",
    primaryClass = "gr-qc",
    doi = "10.1103/PhysRevD.103.104038",
    journal = "Phys. Rev. D",
    volume = "103",
    number = "10",
    pages = "104038",
    year = "2021"
}

@article{Zhang:2025vyx,
    author = "Zhang, Zhenyu and Hou, Yehui and Guo, Minyong and Mizuno, Yosuke and Chen, Bin",
    title = "{Autocorrelation signatures in time-resolved black hole flare images: Secondary peaks and convergence structure}",
    eprint = "2503.17200",
    archivePrefix = "arXiv",
    primaryClass = "astro-ph.HE",
    doi = "10.1103/zmnz-p2rs",
    journal = "Phys. Rev. D",
    volume = "112",
    number = "8",
    pages = "083024",
    year = "2025"
}

@article{Lim:2024mkb,
    author = "Lim, Yen-Kheng and Yeo, Zhi Cheng",
    title = "{Energies and angular momenta of periodic Schwarzschild geodesics}",
    eprint = "2401.13894",
    archivePrefix = "arXiv",
    primaryClass = "gr-qc",
    doi = "10.1103/PhysRevD.109.024037",
    journal = "Phys. Rev. D",
    volume = "109",
    number = "2",
    pages = "024037",
    year = "2024"
}

@article{Huang:2024wpj,
    author = "Huang, Jiewei and Zhang, Zhenyu and Guo, Minyong and Chen, Bin",
    title = "{Images and flares of geodesic hot spots around a Kerr black hole}",
    eprint = "2402.16293",
    archivePrefix = "arXiv",
    primaryClass = "gr-qc",
    doi = "10.1103/PhysRevD.109.124062",
    journal = "Phys. Rev. D",
    volume = "109",
    number = "12",
    pages = "124062",
    year = "2024"
}

@article{Tan:2026yjg,
    author = "Tan, Shijie and Jiang, Chunhua and Li, Dan and Hu, Shiyang and Deng, Chen and Lin, Wenbin",
    title = "{Gravitational emissions and light curves of quasi-periodic orbits in Schwarzschild spacetime embedded in a Dehnen-type dark matter halo}",
    eprint = "2604.13832",
    archivePrefix = "arXiv",
    primaryClass = "gr-qc",
    doi = "10.1016/j.jheap.2026.100685",
    journal = "JHEAp",
    volume = "54",
    pages = "100685",
    year = "2026"
}

@article{Broderick:2005,
    author = "Broderick, Avery E. and Loeb, Abraham",
    title = "{Imaging Bright Spots in the Accretion Flow near the Black Hole Horizon of Sgr A*}",
    doi = "10.1111/j.1365-2966.2005.09458.x",
    journal = "Mon. Not. Roy. Astron. Soc.",
    volume = "363",
    pages = "353--362",
    year = "2005"
}

@article{Luminet:1979,
    author = "Luminet, Jean-Pierre",
    title = "{Image of a Spherical Black Hole with Thin Accretion Disk}",
    journal = "Astron. Astrophys.",
    volume = "75",
    pages = "228--235",
    year = "1979"
}

@article{Levin:2009,
    author = "Levin, Janna and Perez-Giz, Gabe",
    title = "{Homoclinic Orbits around Spinning Black Holes. I. Exact Solution for the Kerr Separatrix}",
    eprint = "0811.3814",
    archivePrefix = "arXiv",
    primaryClass = "gr-qc",
    doi = "10.1103/PhysRevD.79.124013",
    journal = "Phys. Rev. D",
    volume = "79",
    pages = "124013",
    year = "2009"
}

@article{Grossman:2012,
    author = "Grossman, Rebecca and Levin, Janna and Perez-Giz, Gabe",
    title = "{Harmonic Structure of Generic Kerr Orbits}",
    eprint = "1105.5811",
    archivePrefix = "arXiv",
    primaryClass = "gr-qc",
    doi = "10.1103/PhysRevD.85.023012",
    journal = "Phys. Rev. D",
    volume = "85",
    pages = "023012",
    year = "2012"
}

@article{EventHorizonTelescope:2019dse,
    author = "Akiyama, Kazunori and others",
    collaboration = "Event Horizon Telescope",
    title = "{First M87 Event Horizon Telescope Results. I. The Shadow of the Supermassive Black Hole}",
    eprint = "1906.11238",
    archivePrefix = "arXiv",
    primaryClass = "astro-ph.GA",
    doi = "10.3847/2041-8213/ab0ec7",
    journal = "Astrophys. J. Lett.",
    volume = "875",
    pages = "L1",
    year = "2019"
}

@article{EventHorizonTelescope:2019ggy,
    author = "Akiyama, Kazunori and others",
    collaboration = "Event Horizon Telescope",
    title = "{First M87 Event Horizon Telescope Results. VI. The Shadow and Mass of the Central Black Hole}",
    eprint = "1906.11243",
    archivePrefix = "arXiv",
    primaryClass = "astro-ph.GA",
    doi = "10.3847/2041-8213/ab1141",
    journal = "Astrophys. J. Lett.",
    volume = "875",
    number = "1",
    pages = "L6",
    year = "2019"
}

@article{EventHorizonTelescope:2022wkp,
    author = "Akiyama, Kazunori and others",
    collaboration = "Event Horizon Telescope",
    title = "{First Sagittarius A* Event Horizon Telescope Results. I. The Shadow of the Supermassive Black Hole in the Center of the Milky Way}",
    eprint = "2311.08680",
    archivePrefix = "arXiv",
    primaryClass = "astro-ph.HE",
    doi = "10.3847/2041-8213/ac6674",
    journal = "Astrophys. J. Lett.",
    volume = "930",
    number = "2",
    pages = "L12",
    year = "2022"
}

@article{EventHorizonTelescope:2022xqj,
    author = "Akiyama, Kazunori and others",
    collaboration = "Event Horizon Telescope",
    title = "{First Sagittarius A* Event Horizon Telescope Results. VI. Testing the Black Hole Metric}",
    eprint = "2311.09484",
    archivePrefix = "arXiv",
    primaryClass = "astro-ph.HE",
    reportNumber = "FERMILAB-PUB-22-422-PPD",
    doi = "10.3847/2041-8213/ac6756",
    journal = "Astrophys. J. Lett.",
    volume = "930",
    number = "2",
    pages = "L17",
    year = "2022"
}

@article{Gandhi:2008qr,
    author = "Gandhi, P. and Makishima, K. and Durant, M. and Fabian, A. C. and Dhillon, V. S. and Marsh, T. R. and Miller, J. M. and Shahbaz, T. and Spruit, H. C.",
    title = "{Rapid optical and X-ray timing observations of GX 339-4: flux correlations at the onset of a low/hard state}",
    eprint = "0807.1529",
    archivePrefix = "arXiv",
    primaryClass = "astro-ph",
    doi = "10.1111/j.1745-3933.2008.00529.x",
    journal = "Mon. Not. Roy. Astron. Soc.",
    volume = "390",
    pages = "29",
    year = "2008"
}

@article{Fukumura_2010,
   title={QPOs in the time domain: an autocorrelation analysis},
   volume={524},
   ISSN={1432-0746},
   url={http://dx.doi.org/10.1051/0004-6361/201014736},
   DOI={10.1051/0004-6361/201014736},
   journal={Astron. Astrophys.},
   publisher={EDP Sciences},
   author={Fukumura, K. and Shrader, C. R. and Dong, J. W. and Kazanas, D.},
   year={2010},
   month=Nov, pages={A34} }

@article{Conroy:2023kec,
    author = {Conroy, Nicholas S. and Baub{\"o}ck, Michi and Dhruv, Vedant and Lee, Daeyoung and Broderick, Avery E. and Chan, Chi-kwan and Georgiev, Boris and Joshi, Abhishek V. and Prather, Ben and Gammie, Charles F.},
    title = "{Rotation in Event Horizon Telescope Movies}",
    eprint = "2304.03826",
    archivePrefix = "arXiv",
    primaryClass = "astro-ph.HE",
    doi = "10.3847/1538-4357/acd2c8",
    journal = "Astrophys. J.",
    volume = "951",
    number = "1",
    pages = "46",
    year = "2023"
}

@article{Liu:2018vea,
    author = "Liu, Changqing and Ding, Chikun and Jing, Jiliang",
    title = "{Periodic orbits around Kerr Sen black holes}",
    eprint = "1804.05883",
    archivePrefix = "arXiv",
    primaryClass = "gr-qc",
    doi = "10.1088/0253-6102/71/12/1461",
    journal = "Commun. Theor. Phys.",
    volume = "71",
    number = "12",
    pages = "1461",
    year = "2019"
}

@article{Wei:2019zdf,
    author = "Wei, Shao-Wen and Yang, Jie and Liu, Yu-Xiao",
    title = "{Geodesics and periodic orbits in Kehagias-Sfetsos black holes in deformed Hor̆ava-Lifshitz gravity}",
    eprint = "1904.03129",
    archivePrefix = "arXiv",
    primaryClass = "gr-qc",
    doi = "10.1103/PhysRevD.99.104016",
    journal = "Phys. Rev. D",
    volume = "99",
    number = "10",
    pages = "104016",
    year = "2019"
}

@article{Deng:2020hxw,
    author = "Deng, Xue-Mei",
    title = "{Periodic orbits around brane-world black holes}",
    doi = "10.1140/epjc/s10052-020-8067-7",
    journal = "Eur. Phys. J. C",
    volume = "80",
    number = "6",
    pages = "489",
    year = "2020"
}

@article{Zhou:2020zys,
    author = "Zhou, Tian-Yi and Xie, Yi",
    title = "{Precessing and periodic motions around a black-bounce/traversable wormhole}",
    doi = "10.1140/epjc/s10052-020-08661-w",
    journal = "Eur. Phys. J. C",
    volume = "80",
    number = "11",
    pages = "1070",
    year = "2020"
}

@article{Tu:2023xab,
    author = "Tu, Ze-Yi and Zhu, Tao and Wang, Anzhong",
    title = "{Periodic orbits and their gravitational wave radiations in a polymer black hole in loop quantum gravity}",
    eprint = "2304.14160",
    archivePrefix = "arXiv",
    primaryClass = "gr-qc",
    doi = "10.1103/PhysRevD.108.024035",
    journal = "Phys. Rev. D",
    volume = "108",
    number = "2",
    pages = "024035",
    year = "2023"
}

@article{Li:2024tld,
    author = "Li, Yong-Zhuang and Kuang, Xiao-Mei and Sang, Yu",
    title = "{Precessing and periodic timelike orbits and their potential applications in Einsteinian cubic gravity}",
    eprint = "2401.16071",
    archivePrefix = "arXiv",
    primaryClass = "gr-qc",
    doi = "10.1140/epjc/s10052-024-12895-3",
    journal = "Eur. Phys. J. C",
    volume = "84",
    number = "5",
    pages = "529",
    year = "2024"
}

@article{Meng:2024cnq,
    author = "Meng, Liping and Xu, Zhaoyi and Tang, Meirong",
    title = "{Bound orbits and gravitational wave radiation around the hairy black hole}",
    eprint = "2411.01858",
    archivePrefix = "arXiv",
    primaryClass = "gr-qc",
    doi = "10.1140/epjc/s10052-025-14032-0",
    journal = "Eur. Phys. J. C",
    volume = "85",
    number = "3",
    pages = "306",
    year = "2025"
}

@article{Wang:2025hla,
    author = "Wang, Chao-Hui and Meng, Xiang-Cheng and Zhang, Yu-Peng and Zhu, Tao and Wei, Shao-Wen",
    title = "{Equatorial periodic orbits and gravitational waveforms in a black hole free of Cauchy horizon}",
    eprint = "2502.08994",
    archivePrefix = "arXiv",
    primaryClass = "gr-qc",
    doi = "10.1088/1475-7516/2025/07/021",
    journal = "JCAP",
    volume = "07",
    pages = "021",
    year = "2025"
}

@article{Chen:2025aqh,
    author = "Chen, Jiawei and Yang, Jinsong",
    title = "{Periodic orbits and gravitational waveforms in quantum-corrected black hole spacetimes}",
    eprint = "2505.02660",
    archivePrefix = "arXiv",
    primaryClass = "gr-qc",
    doi = "10.1140/epjc/s10052-025-14457-7",
    journal = "Eur. Phys. J. C",
    volume = "85",
    number = "7",
    pages = "726",
    year = "2025"
}

@article{Lu:2025xlp,
    author = "Lu, Shuo and Lin, Hao-Jie and Zhu, Tao and Liu, Yu-Xiao and Zhang, Xin",
    title = "{Gravitational radiations from periodic orbits around a black hole in the effective field theory extension of general relativity}",
    eprint = "2512.11911",
    archivePrefix = "arXiv",
    primaryClass = "gr-qc",
    doi = "10.1140/epjc/s10052-026-15466-w",
    journal = "Eur. Phys. J. C",
    volume = "86",
    number = "3",
    pages = "283",
    year = "2026"
}

@article{Gong:2025mne,
    author = "Gong, Huajie and Long, Sheng and Wang, Xi-Jing and Xia, Zhongwu and Wu, Jian-Pin and Pan, Qiyuan",
    title = "{Gravitational waveforms from periodic orbits around a novel regular black hole}",
    eprint = "2509.23318",
    archivePrefix = "arXiv",
    primaryClass = "gr-qc",
    doi = "10.1140/epjc/s10052-026-15667-3",
    journal = "Eur. Phys. J. C",
    volume = "86",
    number = "5",
    pages = "469",
    year = "2026"
}

@article{Ryan:1995wh,
    author = "Ryan, F. D.",
    title = "{Gravitational waves from the inspiral of a compact object into a massive, axisymmetric body with arbitrary multipole moments}",
    doi = "10.1103/PhysRevD.52.5707",
    journal = "Phys. Rev. D",
    volume = "52",
    pages = "5707--5718",
    year = "1995"
}

@article{Glampedakis:2002ya,
    author = "Glampedakis, Kostas and Kennefick, Daniel",
    title = "{Zoom and whirl: Eccentric equatorial orbits around spinning black holes and their evolution under gravitational radiation reaction}",
    eprint = "gr-qc/0203086",
    archivePrefix = "arXiv",
    doi = "10.1103/PhysRevD.66.044002",
    journal = "Phys. Rev. D",
    volume = "66",
    pages = "044002",
    year = "2002"
}

@article{Bambi:2015kza,
    author = "Bambi, Cosimo",
    title = "{Testing black hole candidates with electromagnetic radiation}",
    eprint = "1509.03884",
    archivePrefix = "arXiv",
    primaryClass = "gr-qc",
    doi = "10.1103/RevModPhys.89.025001",
    journal = "Rev. Mod. Phys.",
    volume = "89",
    number = "2",
    pages = "025001",
    year = "2017"
}

@inbook{Cardenas-Avendano:2024mqp,
    author = "C{\'a}rdenas-Avenda{\~n}o, Alejandro and Sopuerta, Carlos F.",
    title = "{Testing Gravity with Extreme-Mass-Ratio Inspirals}",
    eprint = "2401.08085",
    archivePrefix = "arXiv",
    primaryClass = "gr-qc",
    doi = "10.1007/978-981-97-2871-8_8",
    year = "2024"
}

@article{Misra:2010pu,
    author = "Misra, Vedant and Levin, Janna",
    title = "{Rational Orbits around Charged Black Holes}",
    eprint = "1007.2699",
    archivePrefix = "arXiv",
    primaryClass = "gr-qc",
    doi = "10.1103/PhysRevD.82.083001",
    journal = "Phys. Rev. D",
    volume = "82",
    pages = "083001",
    year = "2010"
}

@article{Zhou:2024dbc,
    author = "Zhou, Lihang and Zhong, Zhen and Chen, Yifan and Cardoso, Vitor",
    title = "{Forward ray tracing and hot spots in Kerr spacetime}",
    eprint = "2408.16049",
    archivePrefix = "arXiv",
    primaryClass = "gr-qc",
    doi = "10.1103/PhysRevD.111.064075",
    journal = "Phys. Rev. D",
    volume = "111",
    number = "6",
    pages = "064075",
    year = "2025"
}

@article{McQuillan_2014,
   title={ROTATION PERIODS OF 34,030
                    <i>KEPLER</i>
                    MAIN-SEQUENCE STARS: THE FULL AUTOCORRELATION SAMPLE},
   volume={211},
   ISSN={1538-4365},
   url={http://dx.doi.org/10.1088/0067-0049/211/2/24},
   DOI={10.1088/0067-0049/211/2/24},
   number={2},
   journal={The Astrophysical Journal Supplement Series},
   publisher={American Astronomical Society},
   author={McQuillan, A. and Mazeh, T. and Aigrain, S.},
   year={2014},
   month=Mar, pages={24} }

@article{Peterson:2004nu,
    author = "Peterson, Bradley M. and others",
    title = "{Central masses and broad-line region sizes of active galactic nuclei. II. A Homogeneous analysis of a large reverberation-mapping database}",
    eprint = "astro-ph/0407299",
    archivePrefix = "arXiv",
    doi = "10.1086/423269",
    journal = "Astrophys. J.",
    volume = "613",
    pages = "682--699",
    year = "2004"
}

@article{Kazanas:1999hg,
    author = "Kazanas, Demosthenes and Hua, Xin-Min",
    title = "{Modeling the time variability of accreting compact sources}",
    eprint = "astro-ph/9902186",
    archivePrefix = "arXiv",
    doi = "10.1086/307379",
    journal = "Astrophys. J.",
    volume = "519",
    pages = "750",
    year = "1999"
}

@article{Berkley:2000mp,
    author = "Berkley, Andrew J. and Kazanas, Demosthenes and Ozik, Jonathan",
    title = "{Modeling the x-ray - uv correlations in ngc 7469}",
    eprint = "astro-ph/0001239",
    archivePrefix = "arXiv",
    doi = "10.1086/308880",
    journal = "Astrophys. J.",
    volume = "535",
    pages = "712",
    year = "2000"
}

@article{Qian:2021aju,
    author = "Qian, Wei-Liang and Lin, Kai and Kuang, Xiao-Mei and Wang, Bin and Yue, Rui-Hong",
    title = "{Quasinormal modes in two-photon autocorrelation and the geometric-optics approximation}",
    eprint = "2109.02844",
    archivePrefix = "arXiv",
    primaryClass = "gr-qc",
    doi = "10.1140/epjc/s10052-022-10155-w",
    journal = "Eur. Phys. J. C",
    volume = "82",
    number = "3",
    pages = "188",
    year = "2022"
}

\end{document}